\documentclass[9pt, onecolumn, aps, pra, showkeys, longbibliography]{revtex4-2}

\usepackage{orcidlink}
\usepackage[T1]{fontenc}
\usepackage[utf8]{inputenc}
\usepackage[english]{babel}
\usepackage{braket}
\usepackage{xcolor}
\usepackage{natbib}
\usepackage{amsfonts, amsmath, amssymb}
\usepackage{graphicx}
\usepackage[top = 2cm, left = 2.5cm, right = 2.5cm, bottom = 2cm]{geometry}
\usepackage{hyperref}
\hypersetup{
    colorlinks = true,
    allcolors = blue,
    pdftitle = {Construction of the Jaynes–Cummings Model over Finite Two-Dimensional Harmonic Oscillators},
}

    \newcommand{\sz}{\hat{\sigma}_{z}}
    \newcommand{\ssp}{\hat{\sigma}_{+}}
    \newcommand{\sm}{\hat{\sigma}_{-}}

    \newcommand{\mi}{\mathrm{i}}

\begin{document}
\title{Construction of the Jaynes–Cummings interaction over the finite two-dimensional oscillator}
\author{Alejandro R. Urz\'ua\,\orcidlink{0000-0002-6255-5453}}
\email{alejandro@icf.unam.mx}
\affiliation{Instituto de Ciencias F\'isicas, Universidad Nacional Aut\'onoma de M\'exico\\ Av. Universidad sn, Col. Chamilpa, 62210, Cuernavaca, Morelos, M\'exico}

\date{\today}

\begin{abstract}
    The interaction between a two-level atom and the finite two-dimensional oscillator in the Cartesian coordinate system is addressed. The construction of the coupling between the degenerate energy states of the finite oscillator and the two levels of the atom allows the identification of an excitation-conserved eigenbasis. A dynamical approach to the evolution of mean values of the fields mediated by the atom is shown in the interaction picture. As an example, the eigenenergies states and $\mathfrak{su}(2)$ coherent states are studied. Perspectives and prospectives are given to settle a path in the study of atomic systems coupled to finite versions of the harmonic oscillator.
\end{abstract}

\keywords{finite oscillators, Jaynes-Cummings model, light-matter interaction, Lie algebras}

\maketitle

\section{Introduction}
    Theoretical studies on the interaction between the quantized radiation field and two-level atomic systems have been addressed since the inception of quantum theory \cite{Heisenberg1925, Von_Neumann2018-ao, Heisenberg1949-en, Dirac1981-ia} and the subsequent branching to quantum optics \cite{Gerry2004-pu}. In this direction, the Jaynes-Cummings model (JCM) \cite{Jaynes1963}, despite its simplicity, has been able to describe fundamental processes between the talking of light and matter. Numerous efforts are underway to utilize the model as a starting point for studying quantum computing, quantum information, and quantum thermodynamics \cite{Bina2012, Azuma2011, Sasaki1996, Oliveira2024, Metwally2004}. In the simplicity of the models lies an elegant potential for pursuing a more profound understanding of one of the most fundamental interactions in nature. As is, the JCM still has a lot to say, and due to its unique algebraic features, possible extensions and generalizations have the potential to provide new insights \cite{Larson2024-wt}. On the algebraic side of the model, we can inquire about the nature of the elements that describe the light field and the two-level atom, where a new realization of the algebras that describe the construction of their spectral properties occurs. One can question the relevance of a model conceived in the mid-20th century, but the evidence suggests that, even sixty years after its initial introduction, the JCM remains relevant today. To support this claim, the last couple of years have seen works pondering the past, present, and future of the model and its potential to continue unraveling details about the quantized light and matter \cite{DeBernardis2024, Knight2024, Larson2024}. To this extent, we can find thousands of results in every scientific research aggregator discussing the JCM, so the vast literature available encourages us to continue asking about what more we can understand about nature through the model. In this manuscript, the focus is on understanding the finite realization of the quantized light field coupled to a two-level atom when this light field has two-dimensional orthogonal components.

    As a direct antecedent and motivation for the finite realization of the quantum harmonic oscillator in the JCM can be found in Ru\'iz, Frank, and Urrutia \cite{MartinRuiz2014}, where they have implemented the finite oscillator to study the standard one-mode Jaynes-Cummings model. Later, they expand its scope to include quantum phase transition, considering this finite field \cite{Quezada2020}. In a clear and evident generalization, the two-mode finite field is in place, which has been extensively studied in its Cartesian \cite{Atakishiyev2001a} and radial \cite{Atakishiyev2001b} coordinate systems by Wolf, Pogosyan, Atakishiyev, and their collaborators in the so-called \emph{Cuernavaca} realization. In this manuscript, we will use the setup given by the Cartesian model. Moreover, there has been interest in extending the study of this two-dimensional finite JCM in the usual (infinite) bosonic algebra; take, for example, the early works of Guo \cite{Gou1989, Gou1990a, Gou1990b}, where he develops a consistent picture of the system. As it stands, the results presented in these works provide a clear picture of what can be achieved in the system in question and how incorporating the two-dimensional finite field will enable us to extend the study of these finite versions of light-matter interaction models. For example, we can define \emph{digital} operations on the finite oscillator fields, like continuous-\emph{like} rotations and gyrations; therefore, Discrete Fractional Fourier Transform (DFrFT) will be available \cite{Urzua2016, Wolf2008, Atakishiyev1999, Urzua2024, Atakishiyev1997}.

    This manuscript is organized as follows: In Section~\ref{finite_osc}, a brief survey about the finite realization of the quantum harmonic oscillator is presented, using the compact $\mathfrak{su}(2)$ algebra. Moreover, finite versions of coherent states are recalled in advance to serve as the initial condition for the wavefield. Next, in Section~\ref{int_jcm}, the detailed construction of the finite two-dimensional oscillator coupled to the two-level atom is presented. In contrast to the standard case of cross-cavity JCM, a discrete and finite version is developed. The Hamiltonian formulation of the system, the eigenbasis, and the general wavefunction are considered to understand the spectral properties and, consequently, the dynamical evolution of average quantities. Section~\ref{dyn_evol} addresses the solution of the Schrödinger equation of the system, trying to show that there exist finite equivalents of well-known quantities like detunings and Rabi frequencies, characteristics of the light-matter interaction. Here, we compare the finite version of the model to a simulation of the cross-cavity JCM, relying on the evolution of mean photonic number values and atomic population. Moreover, it is demonstrated how the system, whether in the finite or infinite algebraic realization, exhibits a fundamental beam splitter-like information sharing between the light fields, mediated by the atomic transitions. Finally, a conclusion is presented in Section~\ref{cons}, summarizing the main results of the work, contrasting perspectives, and advocating a particular perspective for future research on the finite formulation of the oscillatory part of the light-matter interaction.

\section{Revisiting the $\mathfrak{su}(2)$ model of the finite oscillator}\label{finite_osc}

    In an attempt to make the exposition self-contained to our needs, setting apart the fine details that are well documented, we will briefly revisit the formulation of the $\mathfrak{su}(2)$ finite oscillator in terms of the scheme given by Wolf, Atakishiyev, Pogosyan, and collaborators \cite{Arik1999, Alieva2000, Hakioglu2000, ATAKISHIYEV2002}. Let's start with the abstract formulation of the harmonic oscillator. We ask for a set of operators $\{\hat{q}, \hat{p}, \hat{H}\}$, this is position, momentum, and Hamiltonian, respectively. The relations between them, given by the Lie commutator (or Hamilton equations), are
    \begin{equation}
        [\hat{q}, \hat{H}] =: \mi\hat{p},\quad [\hat{p}, \hat{H}] = -\mi\hat{q},
    \end{equation}
    which usually labels the geometric and dynamical postulates. Note that this set of operators closes under a Lie algebra, and by the Jacoby identity, it defines a functional between $\hat{q}$ and $\hat{p}$ of the form
    \begin{equation}
        [\hat{q}, \hat{p}] := \mi\mathcal{G}(\hat{\mathbb{I}}, \hat{H}, \hat{C}),
    \end{equation}
    where $\mathcal{G}$ would depend on the central element, the Hamiltonian, and the Casimir of the algebra, respectively. If we take $\hat{G} = \hat{\mathbb{I}}$, the standard quantum harmonic oscillator is obtained, with algebra $\mathfrak{osc}(1):\{\hat{q}, \hat{p}, \hat{H}, \hat{\mathbb{I}}\}$, containing the Heisenberg-Weyl algebra $\mathfrak{hw}(1):\{\hat{q}, \hat{p}, \hat{\mathbb{I}}\}$ \cite{WOLF1975}. The main features of this oscillator are that it is discrete and semi-infinite in its energy (bounded below), but infinite continuous in its position and momentum \cite{Sakurai2017-sw}. Now, we remember the Lie algebra $\mathfrak{u}(2) = \mathfrak{u}(1)\oplus\mathfrak{su}(2)$. The element $\hat{E}_{j} := j\hat{\mathbb{I}} \in \mathfrak{u}(2)$, sets the label $j$ for the irreducible representations $(2j + 1)$ of $\mathfrak{u}(2)$. On the other hand, the Casimir operator $\hat{C}\in\mathfrak{su}(2)$ will have an eigenvalue spectrum $j(j + 1)$, for $j$ a positive half-integer or integer. In this work we use $j\in\mathbb{Z}_{0}^{+}$.

    In the usual angular momentum theory of $\mathfrak{su}(2)$, the generators of the algebra $\{\hat{J}_{x}, \hat{J}_{y}, \hat{J}_{z}\}$ are related by their Lie commutations
    \begin{equation}
        [\hat{J}_{x}, \hat{J}_{y}] = \mi\hat{J}_{z}\quad\textrm{and antisymmetric permutations},\quad [\hat{J}_{k}, \hat{E}_{j}] = 0,\ k\in\{x,y,z\}.
    \end{equation}
    Here, the identification between the oscillatory set $\{\hat{q}, \hat{p}, \hat{H}\}$ and the angular set $\{\hat{J}_{x}, \hat{J}_{y}, \hat{J}_{z}\}$ could provide a new physical interpretation of the former in terms of the angular momentum generators, this is
    \begin{equation}\label{qp-j}
        \hat{q}\equiv\hat{J}_{x},\quad \hat{p}\equiv-\hat{J}_{y}\quad\textrm{and}\quad \hat{H} = \hat{J}_{z} + (j + \tfrac{1}{2})\mathbb{I}.
    \end{equation}
    In this new interpretation, position, momentum and Hamiltonian has discrete and finite eigenvalues, this are $\Delta_{q}, \Delta_{q} \in \{-j, \cdots, j\}$ and $\Delta_{H} \in \{\tfrac{1}{2}, \cdots 2j + \tfrac{1}{2}\}$. A mode operator is defined as $\hat{H} - \tfrac{1}{2}\mathbb{I} \equiv \hat{J}_{z} + j\mathbb{I}$, with eigenvalues $\Delta_{n}\in\{0, 1, \cdots, 2j\}$, which is commonly denoted $n\equiv\Delta_{n}$, and $2j = N$, the dimension of the space. Due to being $\mathfrak{u}(2)$ a compact and finite space, the equivalences \eqref{qp-j} provide the former oscillatory model with discrete and finite position, momentum, and energy modes, which is patented in the Lie commutation between $\hat{q}$ and $\hat{p}$, this is
    \begin{equation}
        -\mi[\hat{q}, \hat{p}] = \hat{H} - (j + \tfrac{1}{2})\mathbb{I}.
    \end{equation}

    \paragraph*{Energy mode eigenbasis.} Ahead of the presentation of the system Hamiltonian, we will present the eigenbasis from which we will emplace the study. The eigenvalue spectrum $\hat{J}_{z}$ is $\mu\in\{-j, \cdots, j\}$, which is related to the eigenvalue spectrum of the mode energy by $n = \mu + j$ from $\hat{H}$. Between both generators, $\hat{J}_{z}$ and $\hat{H}$, two interrelated eigenstates spawn the whole set of accessible states
    \begin{equation}
        \begin{gathered}
            \hat{J}_{z}\ket{j, \mu}_{z} = \mu\ket{j, \mu}_{z},\quad \mu\vert_{-j}^{j}\qquad \hat{H}\ket{j, n} = n\ket{j, n},\quad n\vert_{0}^{2j}\\
            \vec{J}^{2}\ket{j, \mu}_{z} = j(j + 1)\ket{j, \mu}_{z},\qquad \vec{J}^{2}\ket{j, n} = j(j + 1)\ket{j, n},
        \end{gathered}
    \end{equation}
    where $\ket{j, \mu}_{z} = \ket{j, \mu + j}$. Finally, it is worth mentioning that the finite oscillator starts as a system defined on the $2j + 1$-dimensional position space, in the eigenbasis $\ket{j, q}$, which is related to the $\hat{J}_{1}$ generator. Between these eigenstates in the position set and the energy mode of $\hat{J}_{3}$, a rotation along $\hat{J}_{y}$ relates both representations via
    \begin{equation}
        \underbrace{\ket{j, q}}_{\hat{J}_{1}} = e^{-\mi\tfrac{\pi}{2}\hat{J}_{2}}\underbrace{\ket{j, q + j}}_{\hat{J}_{3}}.
    \end{equation}
    This result implies that the finite oscillator can transition from the configuration or position space to the energy mode space via a transformation kernel given by the Kravchuk functions \cite{Hakioglu2000}. At this time, it is not mandatory to extend the recall on this, as our analysis will begin using the JCM in the energy-related Hamiltonian. But we will keep in mind that recalling the definition and properties of these functions will be needed in subsequent entrances.

    In this one-dimensional model, the $2j + 1$ modes of the oscillator can be reached by the shift generators of the algebra $\hat{J}_{\pm} := \hat{J}_{x} \pm \mi\hat{J}_{y}$, acting on the mode eigenfunctions $\ket{j, \mu}_{z}$ and $\ket{j, n}$ as
    \begin{equation}\label{Jpm}
        \hat{J}_{+}\ket{j, n} = \sqrt{(n + 1)(2j - n)}\ket{j, n + 1},\quad \hat{J}_{-}\ket{j, n} = \sqrt{n(2j - n + 1)}\ket{j, n - 1},
    \end{equation}
    and whose matrix realization are given by arrays of $(2j + 1)\times(2j + 1)$ entries. It is thus that we can generate whatever $n$-state by
    \begin{equation}
        \ket{j, n} = \sqrt{2^{n}\binom{2j}{n}}\ (\hat{J}_{+})^{n}\ket{j, 0}.
    \end{equation}
    
    It is well known that the algebraic path to solve the quantum harmonic oscillator involves the definition of the ladder operators (also known as bosonic operators) $\hat{A}^{\dagger}\propto \hat{q} + \mi\hat{p}$ and $\hat{A}\propto \hat{q} - \mi\hat{p}$, whose realization are infinite matrices, and from which the oscillator algebra $\mathfrak{osc}(1)$ is morphed to $[\hat{A}^{\dagger}, \hat{A}] = 1$, $[\hat{N}, \hat{A}^{\dagger}] = \hat{A}^{\dagger}$ and $[\hat{N}, \hat{A}] = -\hat{A}$, where $\hat{n} := \hat{A}^{\dagger}\hat{A}$ is denoted as the number operator, counting how many excitations the oscillator has in the eigenbasis $\ket{n}$, such that $\hat{n}\ket{n} = n\ket{n}$. This little remembering serves us to recall that it can be shown that the contraction of the ladder generators of the $\mathfrak{su}(2)$ algebra leads, by appropriate scaling, to the well-known ladder generators of the oscillator algebra $\mathfrak{osc}(1)$,
    \begin{equation}
        \lim\limits_{j\rightarrow\infty}\frac{\hat{J}_{+}}{\sqrt{2j}} \rightarrow \hat{A}^{\dagger},\quad \lim\limits_{j\rightarrow\infty}\frac{\hat{J}_{-}}{\sqrt{2j}} \rightarrow \hat{A}.
    \end{equation}

    We will use this contraction property to define a way to map the Jaynes-Cummings Hamiltonian from the infinite realization of the bosonic operators $\hat{A}^{\dagger}$ and $\hat{A}$, to the angular ladder operators $\hat{J}_{\pm}$.
    \begin{equation}\label{AJmap}
        \hat{A}^{\dagger} \triangleq \frac{\hat{J}_{+}}{\sqrt{2j}},\quad \hat{A} \triangleq \frac{\hat{J}_{-}}{\sqrt{2j}}\ \Rightarrow\ [\hat{A}^{\dagger}, \hat{A}] = \hat{\mathbb{I}} - \frac{\hat{n}}{j},\quad \hat{n} \triangleq \hat{J}_{z} + j\mathbb{I}.
    \end{equation}

    \paragraph*{Finite $\mathfrak{su}(2)$ coherent states.} The last recalling in this brief survey section will be on the finite coherent states. There exists a multitude of intents to define the coherent states of the $\mathfrak{su}$(2) finite oscillator; one of the first given by Radcliffe \cite{Radcliffe1971}, W\'odkiewicz and Eberly \cite{Wodkiewicz1985} obtained by the factorization of the exponential map of the algebra. Here, the exposition will be on those coherent states developed in \cite{VICENT2006, Wolf2007}. We can allude to \cite{Uriostegui2024} and references therein for an update on the formulation of coherent states in the finite model. In short, if we took the general eigenstate $\ket{j, \mu}_{z}$ of $\hat{J}_{z}$, the rotation along the $\hat{J}_{y}$ axis will define a collection of eigenstates, which can be obtained as well in the eigenstate $\ket{j, n}$ of $\hat{H}$. Since they are related via a unitary transformation, we get that a finite coherent state is the following action
    \begin{equation}
        \ket{j, \alpha} = e^{-\alpha\hat{J}_{y}}\ket{j, n},\quad 0\leq\alpha\leq\pi.
    \end{equation}
    Note that if we took $\alpha = 0$ and $n = 0$, we observe the canonical coherent state, which is the extremum of the eigenstates in $\hat{J}_{3}$, annihilated by $\hat{J}_{-}$. When $n = 2j$, we get an anti-coherent state, which is annihilated by $\hat{J}_{+}$, in a clear non-correspondence with the case in $\mathfrak{osc}(1)$, where the energy is discrete but unbounded from above. Finally, the coherent state is thus defined as
    \begin{equation}\label{coh_a}
        \ket{j, \alpha} = \sum_{n = 0}^{2j} p_{n}(\alpha)\ket{j,n},
    \end{equation}
    where $p_{n}(\alpha) = \left(\cos\tfrac{1}{2}\alpha\right)^{2j}\left(\tan\tfrac{1}{2}\alpha\right)^{n}\sqrt{\binom{2j}{n}}$, defining the energy mode distribution
    \begin{equation}
    \begin{aligned}\label{pcoh_a}
        \mathcal{P}_{n}(\alpha) &= \vert p_{n}(\alpha)\vert^{2}\\
        &= \left(\cos\tfrac{1}{2}\alpha\right)^{4j}\left(\tan\tfrac{1}{2}\alpha\right)^{2n}\binom{2j}{n},
    \end{aligned}
    \end{equation}
    which is equivalent to the coefficient given in \cite[Eq.~(8)]{MartinRuiz2014}. If we look at the average of the number operator $\hat{n}$ over these finite coherent states, we obtain
    \begin{equation}
    \begin{aligned}
        \langle\hat{n}\rangle &= \langle \alpha, j\vert\hat{n}\vert j, \alpha \rangle\\
        &= \sum_{n = 0}^{2j} p_{n}(\alpha)^{2} \equiv 2j\left(\sin\tfrac{1}{2}\alpha\right)^{2}.
    \end{aligned}
    \end{equation}
    We can see, by the range of $\alpha$, that the maximum value is reached when $\alpha = \pi$, which means that $\langle\hat{n}\rangle = 2j$, so the number of excitations is bounded above, as was expected. Having revisited the formulation of the one-dimensional finite oscillator and its finite coherent states, we can proceed to construct the JCM over the extension of two one-dimensional finite oscillators.

    \section{Interaction on the finite two-dimensional oscillator}\label{int_jcm}

\begin{figure}[htbp]
        \centering
        \includegraphics[width = \linewidth, keepaspectratio]{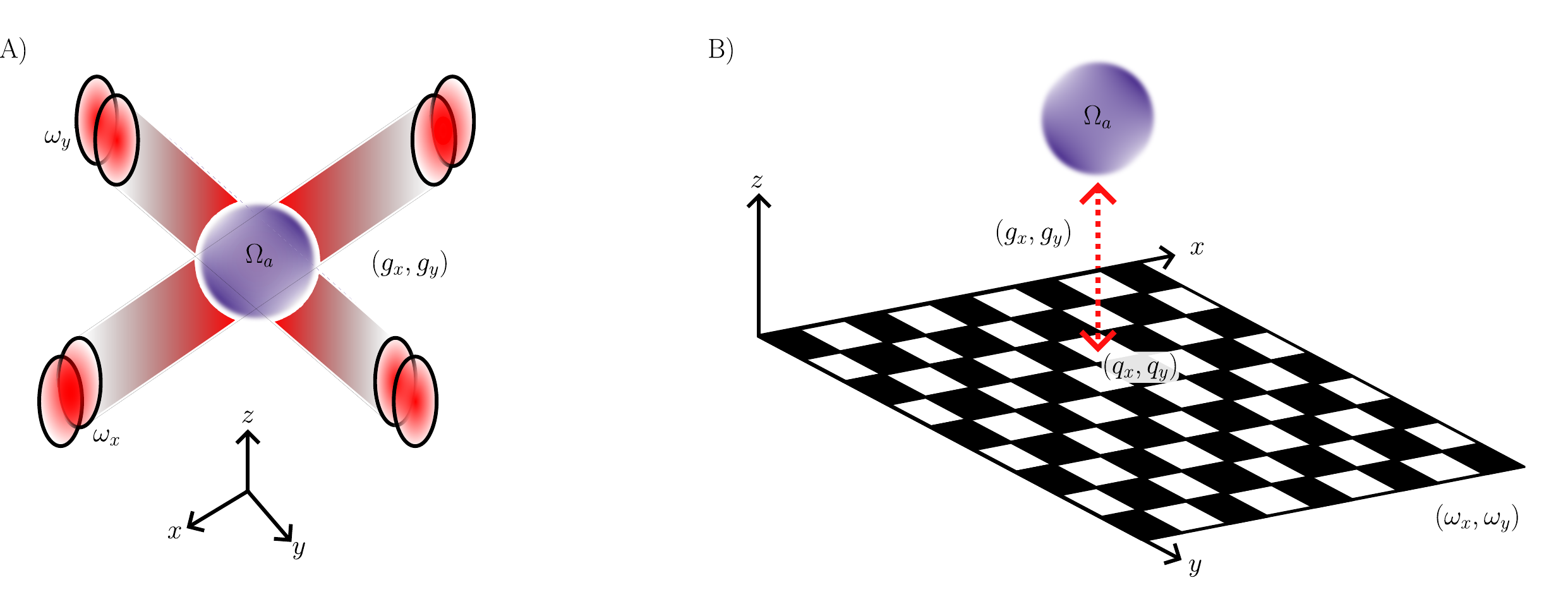}
        \caption{Sketch of the system under consideration. A) A standard cross-cavity setup. An atom with transition frequency $\Omega_{a}$ is placed in the intersection between two lasers with oscillation frequencies $(\omega_{x}, \omega_{y})$, the coupling between the light fields is given by two independent parameters $(g_{x}, g_{y})$. B) A \emph{pixellated screen} representing the spatial modes $\left(q_{x}, q_{y}\right)$ of the two-dimensional finite oscillator, coupled to the atom in front of it, via analogous independent strength couplings $(g_{x}, g_{y})$. This setup will, in principle, mimic the characteristics of the laser fields when a finite description of the available modes is in place.}
        \label{fig:fig_01}
    \end{figure}

    \paragraph*{Schematics of the system. Continuous case.} First, let's describe the continuous version of the system. See Fig.~\ref{fig:fig_01}, panel A). It can be devised as a cross-cavity where two orthogonal lasers confine an atom at the intersection of their paths, as used for para-particles, see Refs.~\cite{Alderete2016, Alderete2018} for example. Each laser has an oscillation frequency $\omega_{x}$ and $\omega_{y}$, respectively; the atom, modeled as a two-level system (TLS), possesses a transition frequency $\Omega_{a}$. In the most general case, the coupling between this two-dimensional oscillator and the two-level atom would be described by the Rabi interaction $\hat{x}_{k}\hat{\sigma}_{x}$, for $k = x\ \text{and}\ y$; this is, the $x$-component of the atomic Pauli matrix and the position quadrature, $\hat{x}_{k} \propto\hat{A}_{k}^{\dagger} + \hat{A}_{k}$, of each oscillator's orthogonal direction. In the spirit of a first approach to this kind of dipolar interaction, we can move to the rotating wave approximation (RWA) with respect to the free evolution of the system, where the atom is coupled to each oscillator by $\hat{A}_{k}^{\dagger}\hat{\sigma}_{-} + \hat{A}_{k}\hat{\sigma}_{+}$. This term refers to the Jaynes-Cummings model of a two-level atom, which is intervened by two distinct oscillators with frequencies and couplings $\left(\omega_{k}, g_{k}\right)$, respectively. We can see the case when $\omega_{x} = \omega_{y} = \omega$ and $g_{x} = g_{y} = g$, where we observe that the rotational symmetry of the system is broken. It is equivalent to having a single Jaynes-Cummings interaction that doubles the oscillatory frequency detuning $\approx\Omega_{a} - \omega_{k}$. In the most general case, conserving the rotational symmetry, we can consider the cross-cavity as a limit case of the finite two-dimensional oscillator, described by the algebra survey we just revisited.

    \paragraph*{Schematics of the system. Finite case.} Armed with the finite description of the one-dimensional oscillator of Section~\ref{finite_osc}, we can scheme the finite case of the two-dimensional system in Cartesian coordinates. See Fig.~\ref{fig:fig_01}. panel B). Consider a square screen where the \emph{pixels} represent the physical positions of the two-dimensional oscillator, labeled by $\left(q_{x}, q_{y}\right)$. In front of this screen, we place an atom with transition frequency $\Omega_{a}$, as in the continuous case. The coupling is given in nature by the equivalent strengths we use for the cross-cavity $\left(g_{x}, g_{y}\right)$. Now, according to the spectral properties of the finite oscillator, there's a connection between the spatial mode $\ket{j, q_{k}}$ and the energy mode $\ket{j, n_{k}}$, and we can transit between both using the Kravchuk functions as the transformation kernel \cite{Wolf2009, Vicent2011}. In the approach of this manuscript, however, we don't need in principle to encode information in the spatial mode, or at least we don't have to deal with it, since the Hamiltonian formulation of the system permits us to work natively in the energy mode $\ket{j, n_{k}}$, which serves as the eigenbasis, along with the atomic modes eigenstates $\left\{\ket{g}, \ket{e}\right\}$. The interaction between this finite two-dimensional version of the cross-cavity, the pixelated screen, and the atom is given by the finite version of the two-mode Jaynes-Cummings Hamiltonian, depicted below.

    \paragraph*{Hamiltonian formulation.} Let's start formulating the setup of a two-dimensional quantum oscillator, and the two-level atom. At first, the oscillatory system consists of two non-interacting fields $\omega_{x}\hat{A}_{x}^{\dagger}\hat{A}_{x}$ and $\omega_{y}\hat{A}_{y}^{\dagger}\hat{A}_{y}$, corresponding to the bosonic algebra $\mathfrak{osc}(1)$. The interaction between them is only possible due to the existence of a two-level atom $\tfrac{\Omega_{a}}{2}\hat{\sigma}_{z}$ (whose algebraic description is in turn given by its own $\mathfrak{su}(2)$ algebra). Refer to the setup depicted in Fig.~\ref{fig:fig_01}, panel A). The Hamiltonian describing this arrangement can be written as
    \begin{equation}
        \begin{aligned}
            \hat{H}_{\texttt{2JC}} &= \hat{H}_{\texttt{osc}}^{(x)} + \hat{H}_{\texttt{osc}}^{(y)} + \hat{H}_{\texttt{atom}} + \hat{V}_{\texttt{int}}\\
            &= \hat{H}_{0} + \hat{V}_{\texttt{int}}
        \end{aligned}
    \end{equation}
    with $\hat{H}_{0} \equiv \hat{H}_{\texttt{osc}}^{(x)} + \hat{H}_{\texttt{osc}}^{(y)} + \hat{H}_{\texttt{atom}}$, and where each of the single infinite oscillator Hamiltonian subsystems is defined by
    \begin{equation}\label{Hosc}
    \begin{gathered}
        \hat{H}_{\texttt{osc}}^{(k)} = \omega_{k}\hat{A}_{k}^{\dagger}\hat{A}_{k},\quad \hat{H}_{\texttt{atom}} = \frac{\Omega_{a}}{2}\sz\\
        \hat{V}_{\texttt{int}} = \sum\limits_{k\in\left\{x, y\right\}}\ g_{k}\left(\hat{A}_{k}^{\dagger}\sm + \hat{A}_{k}\ssp\right).
    \end{gathered}
    \end{equation}
    
    In the finite case of the two-mode Jaynes-Cummings Hamiltonian, we use the finite formulation of the oscillator given above, in Section~\ref{finite_osc}. In the aim to translate the infinite elements of the algebra $\mathfrak{osc}(1)$ to the finite elements of $\mathfrak{su}(2)$ in the oscillator subsystem, remember the relations between $\hat{A}$'s and $\hat{J}$'s, given by Eqs.~\eqref{AJmap}, we can identify the Hamiltonian in the $\mathfrak{su}(2)$ representation by
    \begin{equation}\label{H2jcD}
        \hat{\mathcal{H}}_{\texttt{2JC}} = \sum\limits_{k\in\left\{x,y\right\}} \frac{\omega_{k}}{2j}\hat{J}_{k+}\hat{J}_{k-} + \frac{\Omega_{a}}{2}\sz + \sum\limits_{k\in\left\{x, y\right\}} \frac{g_{k}}{\sqrt{2j}}\left(\hat{J}_{k+}\sm + \hat{J}_{k-}\ssp\right),
    \end{equation}
    which is now a interaction between $\mathfrak{su}(2)\otimes\mathfrak{su}(2)$ algebras. That is, the infinite dimensional elements $\hat{A}$'s of $\mathfrak{osc}(1)$ are replaced by the finite elements $\hat{J}$'s of $\mathfrak{su}(2)$, which is equivalent to the interaction of angular momentum systems by $\hat{J}_{k\pm}\hat{\sigma}_{\mp}$. 

    \begin{figure}[htbp]
        \centering
        \includegraphics[width = 0.75\linewidth, keepaspectratio]{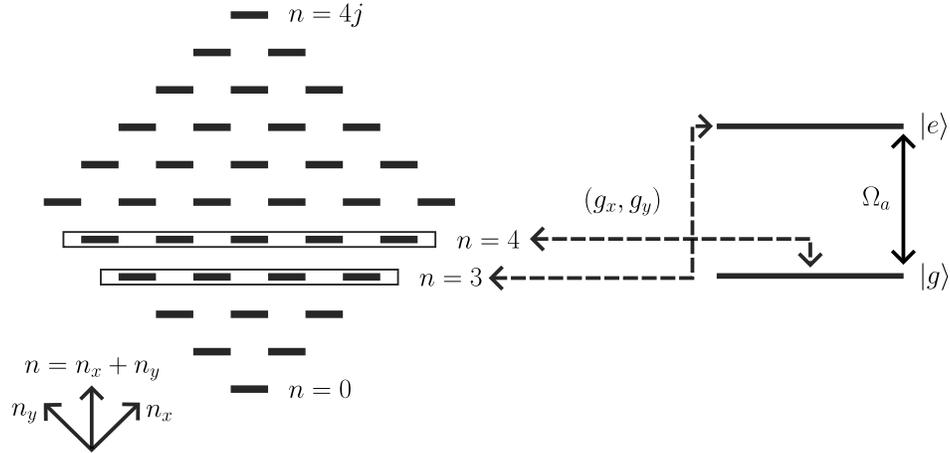}
        \caption{Coupling between finite oscillator states and atomic levels. The $4j$ level states of the finite oscillator, via the conserved excitation operator $\hat{N}$, are coupled to both the ground and excited states of the atom, $\ket{g\textrm{ or }e}$, breaking the inherent degeneracy of the free two-dimensional finite oscillator. To conserve the number of excitations $n + 1$, finite oscillator state $\ket{j, n_{x}, n_{y}}$ is coupled to the excited atom state $\ket{e}$, and the finite oscillator states $\left\{\ket{j, n_{x} + 1, n_{y}}, \ket{j, n_{x}, n_{y} + 1}\right\}$ are coupled to the ground atom state, $\ket{g}$.} 
        \label{fig:fig_02}
    \end{figure}
    
    \paragraph*{Coupled eigenbasis.} The Hamiltonian has a set of uncoupled eigensates, $\ket{j, n_{k}}$ for each finite oscillator, and those for the atom, $\ket{g\textrm{ or }e}$, which define a coupled eigenstate as
    \begin{equation}
        \underbrace{\left\{\ket{j, n_{x}},\ket{j, n_{y}},\ket{g}\textrm{ or }\ket{e}\right\}}_{\mathrm{uncoupled}} \rightarrow \underbrace{\ket{j, n_{x}, n_{y}, g\textrm{ or }e}}_{\mathrm{coupled}},
    \end{equation}
    where the index $j$ defines the size of the representation. Note that these states are acceded by the shift operators $J_{k\pm}$, according to \eqref{Jpm}. As in the infinite case of $\mathfrak{osc}(1)$, the system described by the Hamiltonian \eqref{H2jcD} has a conserved quantity in the number of excitations shared by the eigenmodes between the two finite oscillators and the atom, this is
    \begin{equation}
    \begin{aligned}
        \hat{\mathcal{N}} &= \hat{n}_{x} + \hat{n}_{y} + \frac{1}{2}\left(1 + \hat{\sigma}_{z}\right)\\
                &= \hat{n} + \frac{1}{2}\left(1 + \hat{\sigma}_{z}\right),
    \end{aligned}
    \end{equation}
    which will help us understand how the states in their eigenbasis couple. In Figure~\ref{fig:fig_02}, a schematic way to visualize the $4j$ mode levels of the finite oscillator, $2j$ modes for each, and the two-levels of the atom is shown. Due the existence of a single operator $\hat{n}$, counting the upward oscillator modes, and belonging to the conserved quantity $\hat{N}$, it is coupled to the atom modes via the conserved excitation quantity in the coupled eigenstates $\ket{j, n_{x}, n_{y}, g\textrm{ or }e}$ such that
    \begin{equation}
        \hat{N}\ket{j, n_{x} + l, n_{y} + m, g\textrm{ or }e} \leftrightarrow (n + 1)\ket{j, n_{x} + l, n_{y} + m, g\textrm{ or }e},
    \end{equation}
    where $(l, m)$ are shift indexes. For the set $(l, m) = \{(0,0), (1,0), (0,1)\}$ this unfold the set of eigenstates that shares the same $(n + 1) \equiv (n_{x} + n_{y} + 1)$ coupled excitations, these are
    \begin{equation}\label{eig_span}
        \mathrm{span}\left\{\ket{j, n_{x}, n_{y}, e}, \ket{j, n_{x} + 1, n_{y}, g}, \ket{j, n_{x}, n_{y} + 1, g}\right\},
    \end{equation}
    from which we define the dressed states that belong to the Hamiltonian \eqref{H2jcD}. See Fig.~\ref{fig:fig_02}, as an example, the oscillator level $n = 3$ is coupled to the excited atom level $\ket{e}$, and the oscillator level $n = 4$ is coupled to the ground atom level $\ket{g}$, both observing $n + 1$ excitations, according to $\hat{N}$. This set of coupled eigenstates is the first result presented in this manuscript, as it serves as the spectral construct to define a common eigenbasis, from which we can now study its dynamical properties, for example.
    
    \section{Dynamical evolution}\label{dyn_evol}
        One of the objectives of the manuscript, which defines the way the two-dimensional finite oscillator couples to the atom via the conserving excitation operator, was addressed in the preceding section. It is turn to study, at first stance, this coupling via the dynamical evolution of a suitable initial wavefunction $\ket{\psi(t)}$ obeying the usual Schr\"odinger equation $\mi\partial_{t}\ket{\psi(t)} = \hat{\mathcal{H}}_{\texttt{2JC}}\ket{\psi(t)}$. We can move to a more appealing reference frame; for this, we use the interaction picture of the evolution
    \begin{equation}
        \hat{\mathcal{V}}_{\texttt{int}}(t) = \hat{T}^{\dagger}\hat{V}_{\texttt{int}}\hat{T},
    \end{equation}
    where $\hat{T} = \sum_{k\in\left\{x, y\right\}}\tfrac{\omega_{k}}{2j}\hat{J}_{k+}\hat{J}_{k-} + \tfrac{\Omega_{a}}{2}\hat{\sigma}_{z}$, this is the free part of the Hamiltonian \eqref{H2jcD}. Having performed the bilateral action, we arrive at the expression
    \begin{equation}
        \hat{\mathcal{V}}_{\texttt{int}}(t) = \sum\limits_{k\in\left\{x, y\right\}}\frac{g_{k}}{\sqrt{2j}}\left(
                \hat{S}_{k+}\sm e^{-\mi\Delta(\hat{n}_{k})t} + e^{\mi\Delta(\hat{n}_{k})t}\ssp\hat{S}_{k-}
            \right),
    \end{equation}
    where $\Delta(\hat{n}_{k}) = \tfrac{\Omega_{a}}{2} - \omega_{k}\left(\mathbb{I} - \tfrac{\hat{n}_{k}}{j}\right)$ acts as the finite version of the detuning frequency between each oscillator and the two-level atom, and which limit return the well-known expression $\lim_{j\rightarrow\infty}\Delta(\hat{n}_{k}) = \tfrac{\Omega_{a}}{2} - \omega_{k}$. Moreover, we can point out that when $g_{x} = g_{y}$ and $\omega_{x} = \omega_{y}$, this interaction Hamiltonian corresponds to the double of a one-dimensional finite JCM. In this reference frame, the Schr\"odinger equation of the full Hamiltonian is then replaced by the time-evolution law
    \begin{equation}
        \mi\frac{\partial\ket{\psi(t)}}{\partial t} = \hat{\mathcal{V}}_{\texttt{int}}\ket{\psi(t)},
    \end{equation}
    where the wavefunction $\ket{\psi(t)}$ is to be constructed ahead. Remembering that the eigenspace is spawned by the set of eigenstates in \eqref{eig_span}, we can deduce that the most general time-dependent wavefunction will be
    \begin{equation}
        \ket{\psi(t)} = \sum\limits_{n_{x} + n_{y} = n}^{4j}\Big(
            \mathcal{A}_{n_{x}, n_{y}}(t)\ket{j, n_{x}, n_{y}, e} + \mathcal{B}_{n_{x}, n_{y}}(t)\ket{j, n_{x} + 1, n_{y}, g} + \mathcal{C}_{n_{x}, n_{y}}(t)\ket{j, n_{x}, n_{y} + 1, g}
            \Big),
    \end{equation}
    where $\{\mathcal{A}_{n_{x}, n_{y}}(t), \mathcal{B}_{n_{x}, n_{y}}(t), \mathcal{C}_{n_{x}, n_{y}}(t)\}$ are time-dependent coefficients determined by the initial conditions in the finite fields. Note that under de triangle triangle condition $n_{x} + n_{y} = n$, it does exist boundary conditions, meaning that states like $\ket{j, n_{x} - 2, n_{y}, g}$ are out of the eigenspace, and the legal couplings given by the scheme in Fig.~\ref{fig:fig_02}. These coefficients will be determined by the action of $\mathcal{V}_{\texttt{int}}$ on the wavefunction $\ket{\psi(t)}$, giving the set of nonlinear first-order coupled differential equations
    \begin{equation}\label{diff_eqs}
        \begin{aligned}
            \mi\dot{\mathcal{A}}_{n_{x}, n_{y}}(t) &= g_{x}\sqrt{(n_{x} + 1)\left(1 - \frac{n_{x}}{2j}\right)}\ e^{\mi\Delta(\hat{n}_{x})t}\mathcal{B}_{n_{x}, n_{y}}(t) + g_{y}\sqrt{(n_{y} + 1)\left(1 - \frac{n_{y}}{2j}\right)}\ e^{\mi\Delta(\hat{n}_{y})t}\mathcal{C}_{n_{x}, n_{y}}(t)\\
            \mi\dot{\mathcal{B}}_{n_{x}, n_{y}}(t) &= g_{x}\sqrt{(n_{x} + 1)\left(1 - \frac{n_{x}}{2j}\right)}\ e^{-\mi\Delta(\hat{n}_{x})t}\mathcal{A}_{n_{x}, n_{y}}(t)\\
            \mi\dot{\mathcal{C}}_{n_{x}, n_{y}}(t) &= g_{y}\sqrt{(n_{y} + 1)\left(1 - \frac{n_{y}}{2j}\right)}\ e^{-\mi\Delta(\hat{n}_{y})t}\mathcal{A}_{n_{x}, n_{y}}(t),
        \end{aligned}
    \end{equation}
    subjected to the initial conditions $\{\mathcal{A}_{n_{x}, n_{y}}(0), \mathcal{B}_{n_{x}, n_{y}}(0), \mathcal{C}_{n_{x}, n_{y}}(0)\} = \{\mathcal{A}_{0}, \mathcal{B}_{0}, \mathcal{C}_{0}\}$, determining the initial distributions of excitations in each finite oscillator.

    \paragraph*{Perfect resonant case between oscillators.} We can observe that in the case $g_{x} = g_{y} = g$, and $\omega_{x} = \omega_{y} = \omega \Rightarrow \Delta(\hat{n}_{x}) = \Delta(\hat{n}_{y}) = \Delta(\hat{n})$, a single differential equation for $\mathcal{A}_{n}(t)$ is obtained, such that
    \begin{equation}
        \ddot{\mathcal{A}}_{n}(t) - \mi \Delta(n)\dot{\mathcal{A}}_{n}(t) + 2g^{2}(n + 1)\left(1 - \frac{n}{2j}\right)\mathcal{A}_{n}(t) = 0,
    \end{equation}
    which will return the generalized finite Rabi frequency $\Omega_{n}^{2} = \Delta(n)^{2} + 8g^{2}(n + 1)\left(1 - \tfrac{n}{2j}\right)$, where the bare finite Rabi frequency $\left(\Omega_{n}^{(0)}\right)^{2} = g^{2}(n + 1)\left(1 - \tfrac{n}{2j}\right)$ appears with a coefficient $8$, which doubles the case of the one-dimensional finite JCM \cite[Eq.~22]{MartinRuiz2014}. To obtain, either $\mathcal{B}_{n}(t)$ or $\mathcal{C}_{n}(t)$ it is straightforward since they lead to the same structural differential equation $\dot{\mathcal{B}}_{n}(t)\textrm{ or }\dot{\mathcal{C}}_{n}(t) = g\sqrt{(n + 1)\left(1 - \frac{n}{2j}\right)}\ e^{-\mi\Delta(\hat{n})t}\mathcal{A}_{n}(t)$.

    \paragraph*{Non-resonant case between oscillators.} The general situation when $\omega_{x}\neq\omega_{y}$ and $g_{x}\neq g_{y}$ it's more subtle, since the procedure to obtain the differential equation for $\mathcal{A}_{n_{x}, n_{y}}(t)$ results in a third order expression,
    \begin{equation}
    \begin{gathered}
        \dddot{\mathcal{A}}_{n_{x}, n_{y}}(t) - \mi\left[\Delta(\hat{n}_{x}) + \Delta(\hat{n}_{y})\right]\ddot{\mathcal{A}}_{n_{x}, n_{y}}(t) + \left[\sum\limits_{k\in\{x,y\}}g_{k}^{2}(n_{k} + 1)\left(1 - \frac{n_{k}}{2j}\right) - \Delta(\hat{n}_{x})\Delta(\hat{n}_{y})\right]\dot{\mathcal{A}}_{n_{x}, n_{y}}(t)\\ + \left[\sum\limits_{k\neq l\in\{x, y\}}g_{k}^{2}(n_{k} + 1)\left(1 - \frac{n_{k}}{2j}\right)\Delta(\hat{n}_{l})\right]\mathcal{A}_{n_{x}, n_{y}}(t) = 0,
    \end{gathered}
    \end{equation}
    where the emergent generalized finite Rabi frequency is the roots of the third-order polynomial defined by the differential equation per se, and whose limit in the perfect resonant case returns the result discussed above. It is thus clear, given the nature of this ordinary differential equation, that a closed but cumbersome expression can be obtained. The expressions for the other two coefficients are also obtained by their own differential equations in \eqref{diff_eqs}. However, we need to point it out that the calculation complexity has order $\mathcal{O}(N^{2})$, since there are $N$ operations on each of the finite oscillator and a global sum reigned by $n_{x} + n_{y} = n$, thus we expect that the computational time will increase as the quadratic of $N = 2j$ for each time step $t_{k}$. For $N\rightarrow\infty$, we are close to recovering the description of the usual bosonic $\mathfrak{osc}(1)$ oscillator. Still, for representation sizes $j$ that are large enough, the computation cost will be prohibitive in some cases. Take, for example, the case of $j = 100$ and a time span of $1000$ points. We end up having $1000$ operations for each of the time steps, which is $1000^{2} = 1\times 10^{6}$ operations at the end of the computational routine. 

    \subsection{Expectation values}

    \begin{figure}[b]
        \centering
        \includegraphics[width = \linewidth, keepaspectratio]{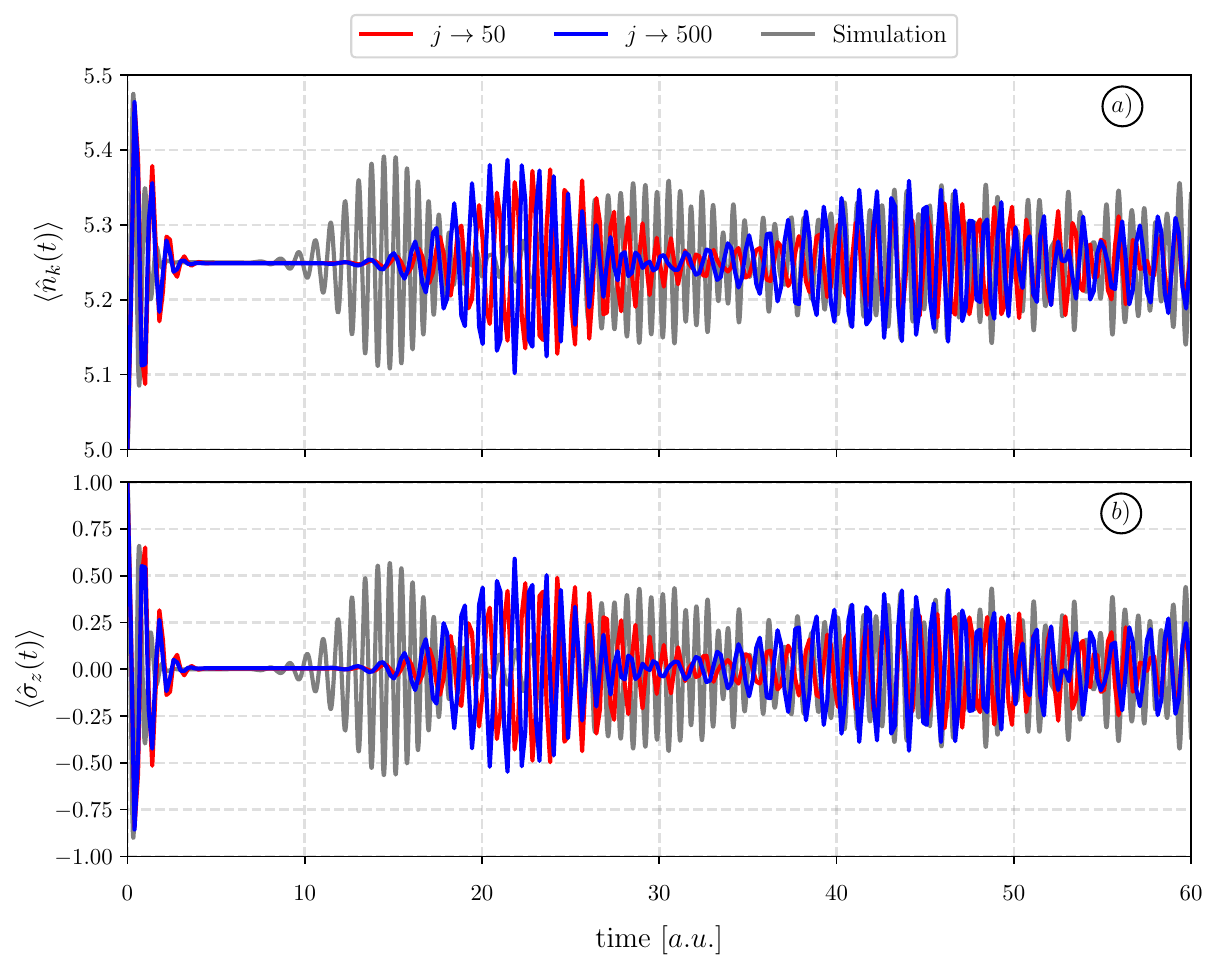}
        \caption{Time evolution of finite coherent states in perfect resonance. Parameters: $\omega_{x} = \omega_{y} = \Omega_{a} = 1.0$, $g_{x} = g_{y} = 1.0$, and initial wavefunction $\ket{\psi(0)} = \ket{j, \alpha_{x}, \alpha_{y}, e}$, $\bar{n}_{x} = \bar{n}_{y} = \sqrt{5}$. In panel a), the mean finite field value for either $\langle\hat{n}_{x}(t)\rangle$ or $\langle\hat{n}_{y}(t)\rangle$. They appear exactly under the perfect resonance condition. In panel b), the atomic population inversion, $\langle\hat{\sigma}_{x}\rangle$. Each finite field version is shifted to the left in the timescale as $j$ increases.}
        \label{fig:fig_03}
    \end{figure}
    
    To quantify the closeness of this finite model to the bosonic model, expectation values for the mean finite fields and the atomic population inversion are calculated. This characterization will enable us to gain confidence in using the model as a playground to implement, in the future, finite oscillator transformations, mode analysis, and information transfer between Cartesian and angular representations. Finite coherent states and finite energy modes (Fock number states) will be studied.
    
    \paragraph*{Coherent states. Perfect resonant case.} Let's first analyze the dynamical evolution of coherent states under this finite treatment. The minimal uncertainty states devised to act as initial wavefunction are described by \eqref{pcoh_a}, and their mean excitation number $\alpha_{k}$ is given by $\vert\alpha_{k}\vert^{2} = \bar{n}_{k}/2j$, where $\bar{n}_{k} = \langle\hat{n}_{k}\rangle$ the expectation to obtain $\bar{n}$ excitations in the current finite coherent state. For the case where $\omega_{x} = \omega_{y} = \Omega_{a}$ and $g_{x} = g_{y}$, we start with $\bar{n}_{k} = 5$ for each finite oscillator, and the atom in its excited level, the initial state is thus $\ket{\psi(0)} = \ket{j, \alpha_{x}, \alpha_{y}, e}$. In Fig.~\ref{fig:fig_03}, the dynamical evolution of the mean finite fields, panels a) $\langle\hat{n}_{x}\rangle$ or $\langle\hat{n}_{y}\rangle$, and the atomic inversion, panel b) $\langle\hat{\sigma}_{z}\rangle$, is shown. A comparison of two values of $j$ is depicted. We can see that, initially, the perfect resonant scenario will cause the evolution of the $x$ and $y$ components of the finite field to appear identical, as expected. The atomic inversion in panel b) indicates that the first revival occurs at different times between the finite and simulated results, despite being on the same timescale. So, there's a slight difference between the $j = 50$ (red line) and $j = 500$ (blue line) cases, denoting convergence to the simulated case (faint gray line); however, this convergence appears to be slow but uniform as $j\rightarrow\infty$.

    \begin{figure}[b]
        \centering
        \includegraphics[width = \linewidth, keepaspectratio]{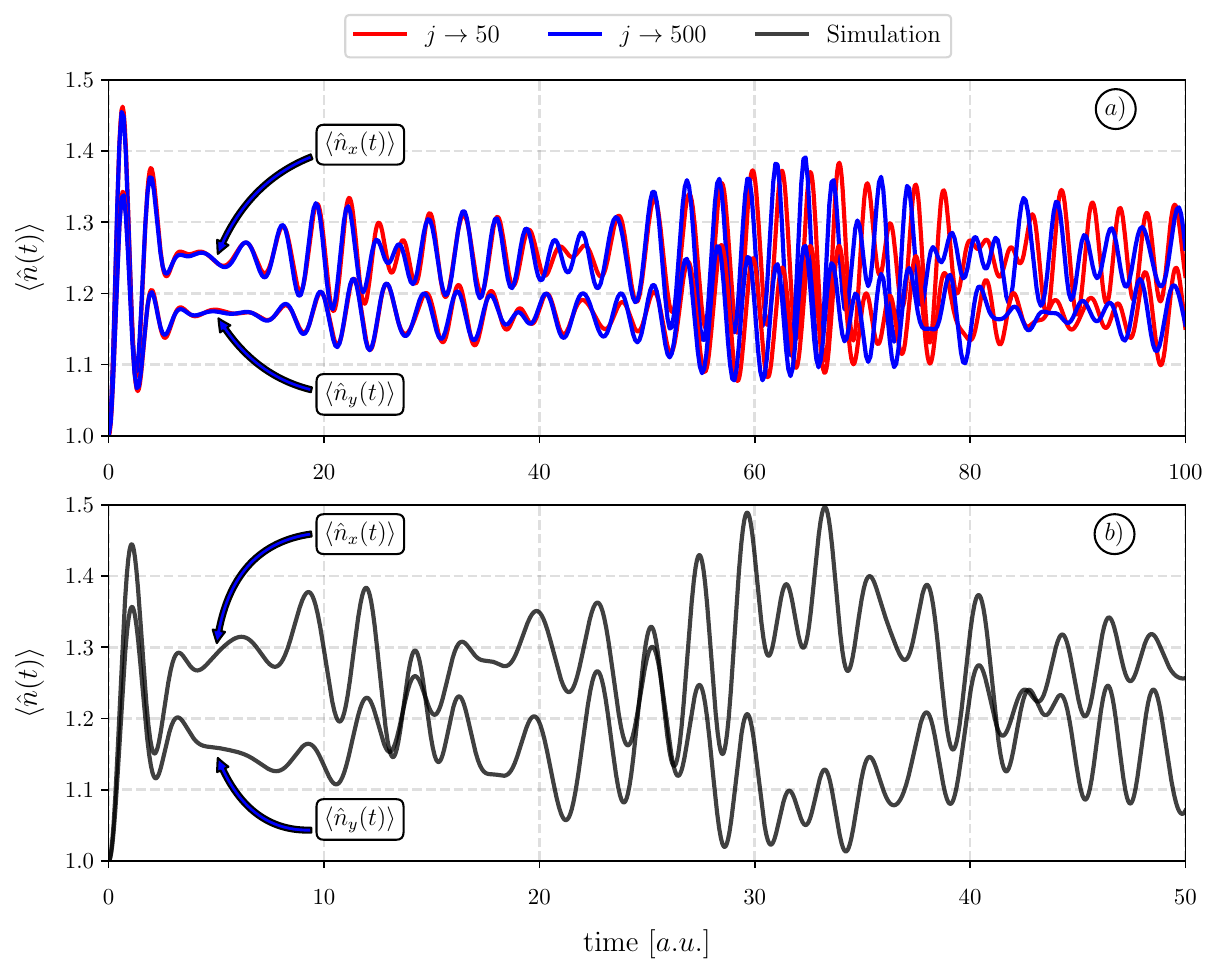}
        \caption{Time evolution of coherent states in non-resonance. Parameters: $\omega_{x} = 1.0$, $\omega_{y} = 0.9$, $\Omega_{a} = 0.1$, $g_{x} = 0.6$, $g_{y} = 0.5$, and initial wavefunction $\ket{\psi(0)} = \ket{j, \alpha_{x}, \alpha_{y}, e}$, $\bar{n}_{x} = \bar{n}_{y} = \sqrt{1}$. In panel a), the splitting between each of the finite fields is shown, compared in panel b) with the splitting given by a simulation. There is a good agreement between the two curves at times proportional to the appearance of the first revival.}
        \label{fig:fig_04}
    \end{figure}

    \begin{figure}[b]
        \centering
        \includegraphics[width = \linewidth, keepaspectratio]{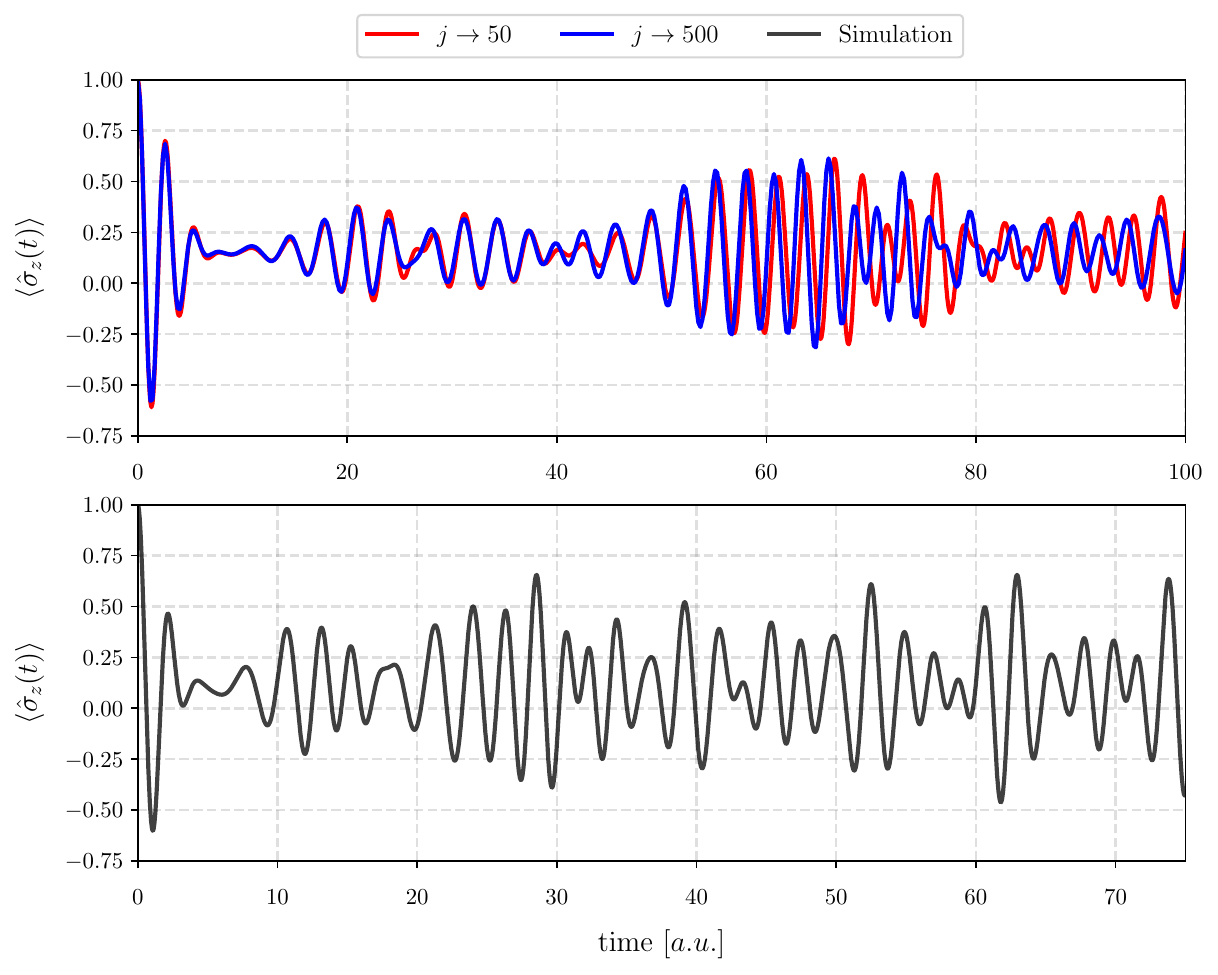}
        \caption{Time evolution of coherent states in non-resonance. Parameters: $\omega_{x} = 1.0$, $\omega_{y} = 0.9$, $\Omega_{a} = 0.1$, $g_{x} = 0.6$, $g_{y} = 0.5$, and initial wavefunction $\ket{\psi(0)} = \ket{j, \alpha_{x}, \alpha_{y}, e}$, $\bar{n}_{x} = \bar{n}_{y} = \sqrt{1}$. In panel a), the atomic population inversion is compared with its analog given by a simulation in panel b). There is a good agreement between the two curves at short times. And again, each finite field evolution is shifted to the left in the timescale as $j$ increases.}
        \label{fig:fig_05}
    \end{figure}

    \paragraph*{Coherent states. Non-resonant case.} Addressing the case where $\omega_{x} \neq \omega_{y} > \Omega_{a}$ and $g_{x} \neq g_{y}$, we would expect to the finite field modes to split, and the atomic inversion to show the influence of this splitting, leading to incomplete atomic population inversion. Here we will take the parameter set: $\omega_{x} = 1.0$, $\omega_{y} = 0.9$, $\omega_{a} = 0.1$, $g_{x} = 0.6$, $g_{y} = 0.5$, and a initial wavefunction $\ket{\psi(0)} = \ket{j, \alpha_{x}, \alpha_{y}, e}$ with $\bar{n}_{x} = \bar{n}_{y} = 5$. Physically, it means that the atom is \emph{slow} compared to the finite fields, transmitting the excitation sharing between the two at a rate slower than the oscillatory field frequencies; however, this case, albeit a little off-resonant, will give us a clear insight into what we would expect in other parameter regimes, far-off detuning and strong or above couplings. First, let's describe the evolution of the mean finite field values in Fig.~\ref{fig:fig_04}. Panel a), shows the time evolution $\langle\hat{n}_{k}\rangle$ for two values of the representation size, $j = 50\textrm{ (red line), } 500\textrm{ (blue line)}$. Panel b), on the other hand, shows the time evolution obtained from simulating the full Hamiltonian. We observe a good agreement between the two depictions, as the splitting between the fields is evident in the finite version. This split between the two modes is due to a term arising in the rotated frame of both field operators, which is detailed in the paragraph below. When we tend the parameters $\omega_{x}\rightarrow\omega_{y}$ and $g_{x}\rightarrow g_{y}$, we recover the results obtained in the section above. Finally, as expected, the finite version is $j$-dependent on the timescale; it shifts the appearance of revivals to the left as this parameter increases. On the other hand, in Fig.~\ref{fig:fig_05}, the time evolution of the atomic population inversion is shown. As it was expected, incomplete inversion is observed within the finite version in panel a), whereas the amplitudes agree with those of the simulation in panel b). These results give us confidence in the finite oscillator development presented in this work. However, we must warn that the strong dependence of the quantum features in the representation size $j$ prevent us to increase arbitrarily the coupling strengths and the oscillation frequencies, or the detuning, since the results will start to diverge from those obtained in the simulation, leading us to conclude that greater $j$ values are required, and computational performance in the same manner.

    \begin{figure}[b]
        \centering
        \includegraphics[width = \linewidth, keepaspectratio]{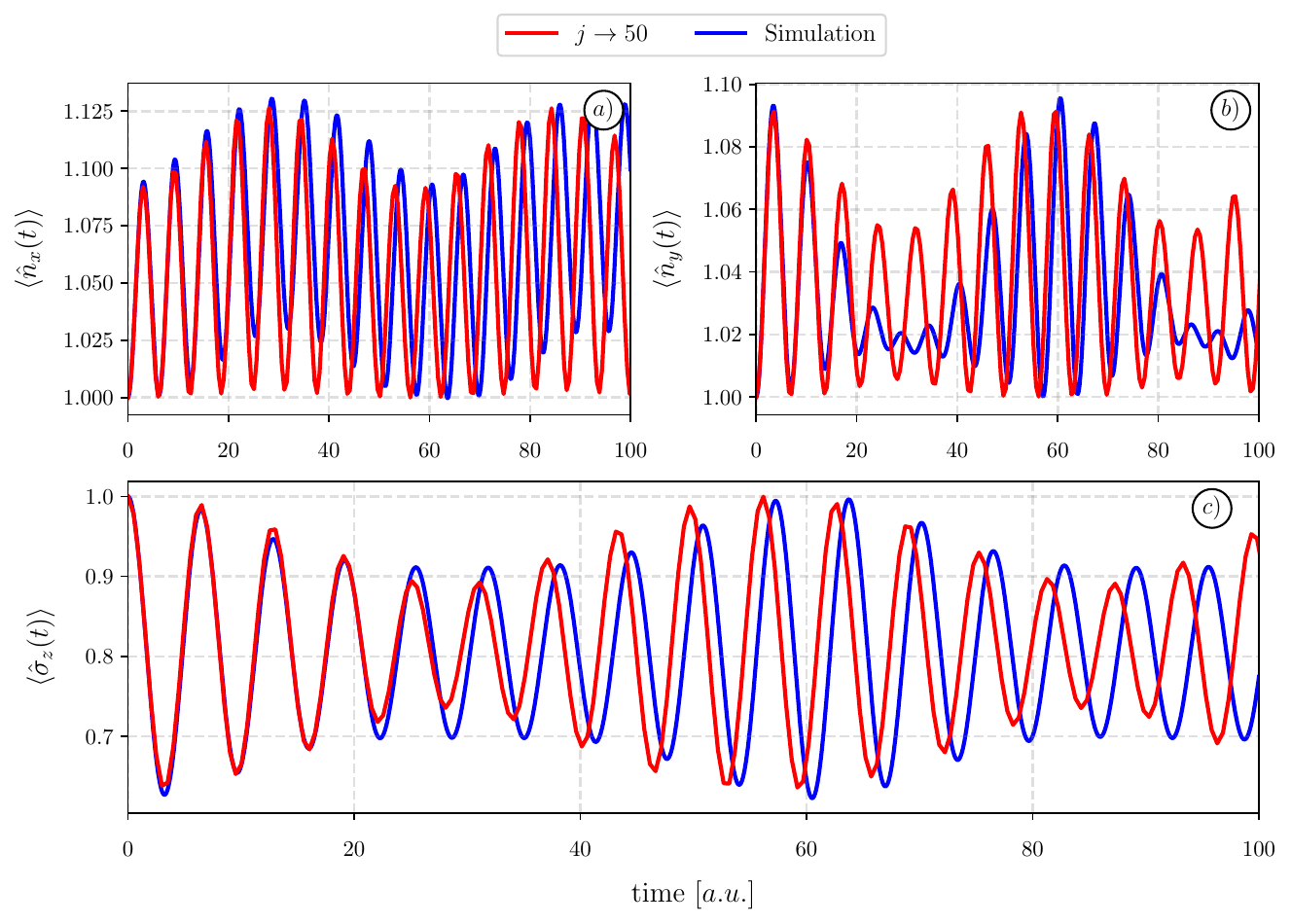}
        \caption{Time evolution of energy mode states in non-resonance. Parameters: $\omega_{x} = 1.0$, $\omega_{y} = 0.9$, $\Omega_{a} = 0.1$, $g_{x} = 0.11$, $g_{y} = 0.10$, and initial wavefunction $\ket{\psi(0)} = \ket{j, n_{x}, n_{y}, e}$, $\bar{n}_{x} = \bar{n}_{y} = 1$. In panels a) and b), the evolution of the finite fields, compared with a simulation, for the unique case $j\rightarrow 50$. In panel c), the atomic population inversion is also compared with a simulation and the same $j$.}
        \label{fig:fig_06}
    \end{figure}

    \paragraph*{Finite energy mode states (Fock number states).} Now we will study the time-evolution of finite energy modes $\ket{j, n}$ of $\hat{J}_{z}$, which are analogous to the Fock number states $\ket{n}$ of $\mathfrak{osc}(1)$. The main difference between the two is their representation, whereas $\ket{n}$ are infinite vectors on Hilbert space, $\ket{j,n}$ are finite-size vectors in $\mathfrak{su}(2)$. However, we can use this set of states as initial wavefunctions to compare their evolution against a simulated version of the original quantum Hamiltonian. Let's then define
    \begin{equation}
        \ket{j,n} = (0, \cdots, \underbrace{1}_{n-\textrm{energy level}},\cdots, 0)^{T},\quad \textrm{dim}(\ket{j,n}) = 2j.
    \end{equation}
    Then we can set the initial condition where finite energy states are present in both finite oscillators, and the atom in its excited state, $\ket{\psi(0)} = \ket{j, n_{x}, n_{y}, e}$. In Fig.~\ref{fig:fig_06} it is shown the time evolution of this initial wavefunction for $\bar{n}_{x} = \bar{n}_{y} = 1$, with fields frequencies $\omega_{x} = 1.0$, and $\omega_{y} = 0.9$, couplings $g_{x} = 0.11$ and $g_{y} = 0.10$, and atomic transition frequency $\Omega_{a} = 0.1$. In panels a) and b), the evolution of the finite fields, compared with a simulation, is shown respectively. It is clear that there is a similarity between the two curves, better in $\langle\hat{n}_{x}(t)\rangle$ than in $\langle\hat{n}_{y}(t)\rangle$; however, as the time span increases, this similarity starts to weaken, this is due to the representation size $j\rightarrow 50$, which is insufficient to track the changes. Nevertheless, we can see good agreement between the finite version and the complete Hamiltonian numerical treatment. In panel c), the atomic population inversion is shown, depicting incomplete inversion in both the finite and the simulation cases. We can see how the finite treatment delivers good agreement between both cases, and starts to diverge in its similitude at times comparable to the appearance of the first quasi-revival or amplitude modulation.
    
    \paragraph*{The beam-splitter sharing between modes.} We can highlight the fundamental interaction between oscillators by looking in the rotated reference frame. For the case, in the bosonic representation of the Hamiltonian \eqref{Hosc}, we apply a unitary transformation of the form
    \begin{equation}
        \hat{R}(\theta) = e^{\theta\left(\hat{A}_{x}^{\dagger}\hat{A}_{y} - \hat{A}_{x}\hat{A}_{y}^{\dagger}\right)}\quad\Rightarrow\quad \begin{cases}
            \hat{R}(\theta)^{\dagger}~\hat{A}_{x}~\hat{R}(\theta) = \hat{A}_{x}\cos\theta - \hat{A}_{y}\sin\theta\\
            \hat{R}(\theta)^{\dagger}~\hat{A}_{y}~\hat{R}(\theta) = \hat{A}_{y}\cos\theta + \hat{A}_{x}\sin\theta
        \end{cases},
    \end{equation}
    from where we obtain the Hamiltonian 
    \begin{equation}
    \begin{aligned}
        \hat{R}(\theta)^{\dagger}\hat{H}_{\texttt{2JC}}\hat{R}(\theta) &= \hat{A}_{x}^{\dagger}\hat{A}_{x}\left(\omega_{x}\cos^{2}\theta + \omega_{y}\sin^{2}\theta\right) + \hat{A}^{\dagger}\hat{A}\left(\omega_{y}\cos^{2}\theta + \omega_{x}\sin^{2}\theta\right) + \left(\hat{A}_{x}^{\dagger}\hat{A}_{y} + \hat{A}_{x}\hat{A}_{y}^{\dagger}\right)\cos\theta\ \sin\theta\\
        &+ \left(\hat{A}_{x}^{\dagger}\hat{\sigma}_{-} + \hat{A}_{x}\hat{\sigma}_{+}\right)\left(g_{x}\cos\theta + g_{y}\sin\theta\right) + \left(\hat{A}_{y}^{\dagger}\hat{\sigma}_{-} + \hat{A}_{y}\hat{\sigma}_{+}\right)\left(g_{y}\cos\theta + g_{x}\sin\theta\right) + \frac{\Omega_{a}}{2}\hat{\sigma}_{z},
    \end{aligned}
    \end{equation}
    in such way that $\theta$ can be choose to banish either the $x$-JCM or the $y$-JCM, but not simultaneous. In contrast, the payback to make this is leave a term proportional to $\hat{A}_{x}^{\dagger}\hat{A}_{y} + \hat{A}_{x}\hat{A}_{y}^{\dagger}$, which is a beam-splitter term, clearly denoting the symmetric interchange of excitations between each mode of the oscillators, through the presence of the atom. In the case we want to conserve the $x$-JCM, we obtain
    \begin{equation}
        \hat{H}_{\texttt{2JC}}^{(R)} = \underbrace{\left[\bar{\omega}_{x}\hat{A}_{x}^{\dagger}\hat{A} + \frac{\Omega_{a}}{2}\hat{\sigma}_{z} + \bar{g}_{x}\left(\hat{A}_{x}^{\dagger}\hat{\sigma}_{-} + \hat{A}_{x}\hat{\sigma}_{+}\right)\right]}_{x\textrm{-JCM}} + \underbrace{\left[\bar{\omega}_{y}\hat{A}_{y} ^{\dagger}\hat{A}_{y} + \bar{g}\left(\hat{A}_{x}^{\dagger}\hat{A}_{y} + \hat{A}_{x}\hat{A}_{y}^{\dagger}\right)\right]}_{y\textrm{-free oscillator and beam splitter}},
    \end{equation}
    for the new coefficients given by $\bar{\omega}_{x} = (\omega_{x}g_{x}^{2} + \omega_{y}g_{y}^{2})/\bar{g}_{x}^{2}$, $\bar{g}_{x} = \sqrt{g_{x}^{2} + g_{y}^{2}}$, $\bar{\omega}_{y} = (\omega_{y}g_{x}^{2} + \omega_{x}g_{y}^{2})/\bar{g}_{x}^{2}$ and $\bar{g} = -2g_{x}g_{y}(\omega_{x} - \omega_{y})/g_{x}^{2}$. A fine-tuning in the rotated frame can be done to devise a system where the beam-splitter term is disregarded. But the principal reason for revisiting this transformed Hamiltonian is to explicitly show that we can expect interchange of excitations between each mode of the oscillator $x$ and $y$. And as for the finite oscillator, the mapping $\hat{A}_{k}\mapsto\hat{J}_{k}$ would be valid under the supposition discussed above.

\section{Conclusions}\label{cons}

In this manuscript, the construction of the Jaynes-Cummings model over the two-dimensional finite oscillator is accomplished. The construction is realized under the Cartesian generalization of the finite oscillator of $\mathfrak{su}(2)$ with representation size $j$ for each of the directions, $j_{x} = j_{y} = j$, thus giving a square joint representation. The coupling between the $4j$ mode energy states $\ket{j, n_{x}, n_{y}}$ of the oscillator to the two-level atomic transition $\ket{g\textrm{ or }e}$ is given, where the existence of an excitation conserved operator $\hat{\mathcal{N}}$ allows us to identify the minimal set of common eigenstates spawning the joint eigenbasis $\ket{j, n_{x}, n_{y}}\otimes\ket{g\textrm{ or }e}$. The general wavefunction $\ket{\psi(t)}$ is defined as a preamble to study the construction under the dynamical evolution of their mean field and atomic values. The time evolution of the system for number states and coherent states was shown, emphasizing the strong dependence of the features with respect to the $\mathfrak{osc}(1)$ case as a function of the representation size $j$. One of the caveats in this proposal, is that of the demanding computational time, since each of the finite oscillators involved needs to know its own $N^{2} = (2j)^{2}$ states, plus a lump sum that will lead to a complexity order $\mathcal{O}(N^{3})$, where for larger $j$ values leads to a prohibitive computational time. Analytically, the limit $j\rightarrow\infty$ is well defined, and each of the finite oscillators in $\mathfrak{su}(2)$ converges to the bosonic oscillator $\mathfrak{osc}(1)$.

Summarizing, the development of this model into a finite representation reveals its value, as it can serve as the starting point for integrating finite-mode analysis, finite-field rotations and gyrations, finite mapping between Cartesian and angular representations of the oscillators, and finite fractional Fourier transforms, among other applications. All in a platform that enjoys the perks of having an atomic sidekick, allowing them to interact with each other, opening new paths to a broader understanding of finite-size oscillators coupled to atomic systems. In this line, the logical next step will be to have a three-dimensional finite oscillator and a two-level atom, where pioneer work has been done defining unitary rotations \cite{Urzua2022}, or the case where each of the finite oscillators has different representation size $j_{x}\neq j_{y}$, leading to a rectangular rather square scheme of interaction \cite{Urzua2016}. Finally, more complex but richer systems, such as the general Rabi and Dicke Hamiltonians, can, in principle, be formulated within the finite scheme.

\pagebreak
\section*{Acknowledgments} A.R.-U. acknowledges financial support by UNAM Postdoctoral Program (POSDOC) 2024-2025, and to ICF-UNAM for the assistance in place. The author is indebted to F. R\'ecamier-Angelini and L. Medina-Dozal for providing the motivation and feedback about the two-dimensional version of the JCM, and an acknowledgement to Reyes Garc\'ia (C\'omputo-ICF) for maintaining the computing servers.

\bibliography{biblio}

\end{document}